\documentclass[epsfig,bm,showpacs,twocolumn, preprintnumbers]{revtex4}
\usepackage{amsmath, amsfonts,euscript,bbm}
\usepackage{ifthen}
\usepackage{graphicx}
\usepackage{psfrag}

\newboolean{Notes}\setboolean{Notes}{true}
\newcommand{\notes}[1]
	    {\ifthenelse{\boolean{Notes}}{{\tt #1}}{}}
\newcommand{\bas}{\begin{eqnarray*}}
\newcommand{\eas}{\end{eqnarray*}}
\newcommand{\ba}{\begin{eqnarray}}
\newcommand{\ea}{\end{eqnarray}}
\def\be{\begin{equation}}
\def\ee{\end{equation}}
\def\ben{\begin{equation} \nonumber}
\def\een{\end{equation}}

\def\h{h_A^{2}} 

\def\hf{\frac12}
\def\ap{{\alpha^{\prime}}}

\def\nalpha{n_\alpha}

\def\Nalpha{N_\alpha}
\def\Nbeta{N_\beta}
\def\Ngamma{N_\gamma}

\begin{document}
\preprint{MIFP-07-01}
\preprint{astro-ph/0701427}

\title{Cosmic Necklaces from String Theory}

\author{Louis Leblond$^1$}
\email{lleblond@physics.tamu.edu}
\author{ Mark Wyman$^2$}
\email{mwyman@perimeterinstitute.ca}
\affiliation{$^1$Texas A\&M University, Mitchell Institute for Fundamental Physics, College Station, Texas, 77840, USA\\
$^2$Perimeter Institute for Theoretical Physics, 31 Caroline St. N, Waterloo, ON, N2L 2Y5, Canada}
\begin{abstract}
We present the properties of a cosmic superstring network in the scenario of flux compactification. An infinite family of strings, the (p,q)-strings, are allowed to exist.  The flux compactification leads to a string tension that is periodic in $p$.   Monopoles, appearing here as beads on a string, are formed in certain interactions in such networks. This allows bare strings to become cosmic necklaces. We study network evolution in this scenario, outlining what conditions are necessary to reach a cosmologically viable scaling solution. We also analyze the physics of the beads on a cosmic necklace, and present general conditions for which they will  be cosmologically safe, leaving the  network's scaling undisturbed. In particular, we find that a large average loop size is sufficient for the beads to be cosmologically safe. Finally, we argue that loop formation will promote a scaling solution for the interbead distance in some situations.
\end{abstract}
\maketitle
\section{Introduction}

Braneworld models of inflation can produce cosmic strings \cite{Jones:2002cv, Sarangi:2002yt, Becker:2005pv}.
This discovery has reignited interest in cosmic strings as cosmological observables (see, e.g., \cite{Tye:2006uv, Polchinski:2004ia} for reviews). Since cosmic strings can also 
be formed in ordinary gauge-theory inflationary models, string theorists are 
eager to find ways in which the cosmic ``superstrings" from brane inflation are different 
from the Abelian Higgs cosmic strings typically found in gauge-theory inflation. 

Among these distinctive phenomena, the best known is that higher-dimensional 
effects can cause the superstring intercommutation probability to be 
considerably less than one  \cite{Jackson:2004zg, Jackson:2006rb};
this reduced probability does not significantly alter
 network dynamics in most cases \cite{Avgoustidis:2005a}.
However, an even more exotic new property of cosmic superstrings is that
 they can come in more than one fundamental species.
The braneworld permits the formations of both fundamental cosmic strings, or F-strings, and Dirichlet brane cosmic strings, or D-strings.  This leads to a new type of string interactions: binding of multiple strings.  
The network-level scaling behavior of these multi-tension, 
binding strings was first studied in Ref. \cite{Tye:2005fn}.  In that work, however,
they assumed a simple string tension formula that is only valid for strings living in
a flat 10 dimensional Minkowski space, which is not a realistic scenario. An important goal
of this paper is to establish conditions under which cosmic superstring networks can reach
a scaling solution when they are realized in a more realistic -- i.e., a phenomenologically
viable --  background geometry.

Constructing more-realistic string theory models of the early universe has led to a great deal of work
on how the complexities of higher-dimensional geometry are tied to physics.
In particular, braneworld models of inflation have been realized in a 
geometry known as the Klebanov-Strassler (KS) throat \cite{Klebanov:2000hb}.
Inspired by these developments, Firouzjahi et al. \cite{Firouzjahi:2006vp} showed that
the properties of cosmic superstrings can depend on the structure of this extra-dimensional geometry.

It has also been discovered that fundamental strings in a KS throat are dual to confining flux tubes. 
Since confining flux tubes are important in many contexts, discovery 
of this relationship has occasioned yet more investigation of these strings.
In Ref. \cite{Herzog:2001fq}, it was found that fundamental 
strings can join on a D3-brane wrapping the non-vanishing $S^3$ 
cycle that lives at the bottom of a KS throat.  Since this is a three-dimensional object being wrapped
on a three-dimensional cycle, the wrapped D3-brane looks like a point charge, or a monopole, in
4D.  In the dual gauge theory, this monopole is equivalent to a baryon. 
Thus, cosmic superstrings networks can also contain monopoles.  Since these monopoles
 are dual to baryons in the gauge theory, one might call these stringy monopoles ``baryons," too.
However, since this would likely cause confusion, we
will instead call these objects ``beads."  These ``beads" are stuck on cosmic strings and will act like beads on the cosmic ``string". Cosmic strings from string theory are thus actually cosmic ``necklaces."
Such string-bead ``necklace" objects were first studied in the context of gauge theory defect models in Ref. \cite{Hindmarsh:1985xc}. Also, a different approach to cosmic necklaces in string theory can be found in Refs. \cite{Matsuda:2005fb, Matsuda:2004bg}.

As it turns out, Fundamental strings living in the warped KS throat geometry
have a tension spectrum that is sinusoidal in its dependence
on winding number, or flux quantum number.
The period is set by the amount of RR $F_3$ flux in the extra-dimensions. This is an integer
denoted M and the fact that the tension is periodic is an
important new property that we must take into account.
In this paper, we will treat $M$ as a free parameter. We will
show that requiring safe cosmological behavior already places lower limits 
on what values $M$ can have. 
To summarize: a collection of  $M$ F-strings can join on a bead. 
They are $Z_M$ charged under the usual 
NS-NS 2-form $B_2$  (which in 4D is Hodge dual to an axion).
These facts, that cosmic F-strings are $Z_M$ charged and can 
join on beads, were first discussed by Ref. \cite{Copeland:2003bj}.

Dirichlet strings, or D-strings, on the other hand, are believed to be 
BPS in the KS throat \cite{Gubser:2004qj}. The F-strings and D-strings together form a $(p,q)$ string network. This network is so named because the mutual interactions between F and D strings are binding interactions. Thus, a series of interactions between $p$ F-strings and $q$ D-strings can result in the formation of a bound state containing the net charge for all the bound strings. This string bound state is called a $(p,q)$ string. The tensions of the various $(p,q)$ states in 
a warped throat were calculated in Ref. \cite{Firouzjahi:2006vp} to be
\begin{equation}\label{finalanswer}
T_{p,q} \simeq  m_s^2
 \sqrt{  \frac{q^2}{g_s^2} + 
\left (\frac{b M}{\pi} \right )^2 \sin^2 \left (\frac{\pi p}{M} \right)},
\end{equation}
where $m_s = \sqrt{\frac{\h}{2 \pi \ap}}$ is the warped string scale, $g_s$ is the string coupling constant and b is a numerical factor approximately equal to $0.93$.  This formula was also derived using different methods in Refs. \cite{Firouzjahi:2006xa, Thomas:2006ud}.

As we noted previously, multi-tension networks of binding $(p,q)$ strings have been studied before \cite{Tye:2005fn, Jackson:2004zg}. One goal of our present work is to generalize the work of Ref. \cite{Tye:2005fn} using the more realistic tension spectrum (Eqn. \ref{finalanswer}) and taking into 
account  periodicity in the quantum number, $p$. 
We also hope to understand better the ``beads" that can be formed dynamically
on these strings. For future reference, the mass of the beads is given by \cite{Gubser:1998fp}
\begin{align}
M_b = \left(\frac{bM}{\pi}\right)^{3/2} \frac{m_s\sqrt{g_s}}{3\sqrt{2}}.
\end{align}

This paper is roughly divided into two different parts. The first is concerned with the scaling of the network 
while neglecting the beads. The second part discusses the physics of the beads themselves.
We study the scaling of the network neglecting the beads using a modified version of the VOS model \cite{Martins:2000cs}.  As this study will closely follow that of Ref. \cite{Tye:2005fn}, we will not provide a detailed review of that model.

The main result for the first part of our paper is that the network does have a scaling solution when $M \gtrsim 10$.
The more complex background geometry also leads to new network features.
For instance, the structure of the tension spectrum is much richer than 
when using the flat space formula for the tension, and the overall 
cosmic string density is slightly higher.  Interestingly, we also find that the scaling
density of cosmic strings depends on the network's initial conditions, with there
being, in effect, two possible scaling solutions.
If the large quantum number string states -- those with $p \sim  M$ -- are excited initially,
then the network will scale with comparable number density of $(1,0)$ and $(M-1,0)$ strings,
effectively doubling the scaling network energy density as compared with the flat space case. We remind
the reader that $(1,0)$ and $(M-1,0)$ strings have the same tension. 
On the other hand, if the post-inflationary network creation excites exclusively
low quantum number strings -- those with $p<M/2$ -- then the network 
will reach scaling at a density only slightly higher than the flat space case. 
Though the former scenario is intriguing, we consider the latter to be more probable.

All of the results summarized here were derived under the assumption that the beads' 
contribution to the network's energy density is negligible. We examine this issue
in detail in section \ref{beadanalysis}.
To ease discussion of this issue, we will follow Berezinsky and Vilenkin \cite{Berezinsky:1997td} by quantifying
the beads' contribution to the energy density with a dimensionless parameter, $r$:
\ba
r = \frac{M_b}{\mu d}
\ea
where $d$ is the average interbead distance along the string and $\mu$ is the tension of the string. The energy per unit of length of a bead-carrying string is, then, $(r+1)\mu$.  For small $r$, the beads contribute a negligible amount to the energy density and can be neglected, while for large $r$ they are cosmologically disastrous:
the network becomes a gas of monopoles connected by relatively light, irrelevant strings.

We first analyze the evolution of $r$ in the scaling regime, neglecting loop formation, annihilation of beads,
and taking the assumption that the 
network reaches scaling with a small value of $r$ after an early, transient regime.
In this case, we find that $r$ grows unless the average size of
loops (which controls the rate of gravitational wave emission and thus string contraction)
is bigger than about 
\ba
l_{\rm{loop}} > 10^2 G\mu t
\ea
where $t$ is the FRW coordinate of time.

Loop formation should, in general, make $r$ increase as well, even if the average loop size respects
the previous bounds. But taking account of loops will not always force $r$ to grow indefinitely.
Instead, we believe $r$ should reach a scaling value, with the interbead distance of the same order as the average loop size.  We argue for this view more extensively in section \ref{loops}. 
In this approach, the approximate scaling value for $r$ should be given by
\ba
 r \sim 10^2 M^{3/2} \frac{\sqrt{2g_s b}}{3\pi^{3/2}} \frac{H}{m_s}
\ea
where $H\sim 1/t$ is the Hubble constant.  
 Strangely, under
these assumptions we find that larger values of the flux, $M$,
lead to larger values of $r$ and, thus, potentially more observable features.  This is a counter-intuitive 
result, since in the large $M$ limit we recover the flat space tension formula, 
which does not share any of these new features. Dynamical bead formation is also heavily suppressed
for large values of $M$. Thus, we should regard this result as suggestive but incomplete.
The proportionality between $r$ and $M$ reflects the fact that monopole mass grows with $M$ more quickly than does
the string tension. An interim regime, where $r\sim1$, would be very interesting to study.
In this case, the effect of beads would cosmologically be safe, yet observable:
essentially doubling the effective tension of every string in the network.  However, even in the most optimistic case, the fact that $r\propto H$ means that any period of bead-string energy balance will be short lived; as $H$ shrinks with cosmological
expansion, $r$ will also shrink.  It is thus unlikely that there will be a great number of necklace-specific
effects that remain important long enough to be easily observable.

Finally we comment on the physics of the beads in the transient regime, when the 
network evolves from its initial configuration to a scaling solution. This is  a much murkier issue. 
 Only a complete simulation can really address this epoch with any certainty.  
 Yet it is possible to get an idea for what may happen during this regime by employing 
the same logic we use in the scaling era. We expect
$r$ to increase significantly during this regime -- during which we expect  a great number of string
interactions to occur -- so that we will need a large value of $M$ in order to prevent $r$ from growing beyond 1.
If we take quantum mechanical tunneling suppression of bead formation as the main
mean for limiting the growth of $r$, we find that $M$ should be $\sim 10^3$ or more to control
the growth of $r$ during the transient regime. But, again, this result should be taken as very conjectural,
 given the complexity of network physics outside of the scaling era.

Each of these issues will be better fleshed out below. In section \ref{interactions}, we summarize the main properties
of the cosmic string network. In section \ref{network}, we discuss the network's scaling solution, obtained
through numerical integration of a modified MTVOS network model. Section \ref{beadanalysis} treats the physics
of the beads in more detail.

\section{Interactions}\label{interactions}

We expect no substantial change from 
the calculation of Ref. \cite{Jackson:2004zg} for local binding events. 
Differences appear only when the interaction forms beads.

\subsection{Probability of Interaction}
\begin{figure}[ht]
\centering
\psfrag{M-1}{$M-1$}
\psfrag{1}{$1$}
\psfrag{M-1}{$M-1$}
\psfrag{1}{$1$}
\psfrag{M-1}{$M-1$}
\psfrag{1}{$1$}
\psfrag{M-1}{$M-1$}
\psfrag{1}{$1$}
\includegraphics[width=8cm]{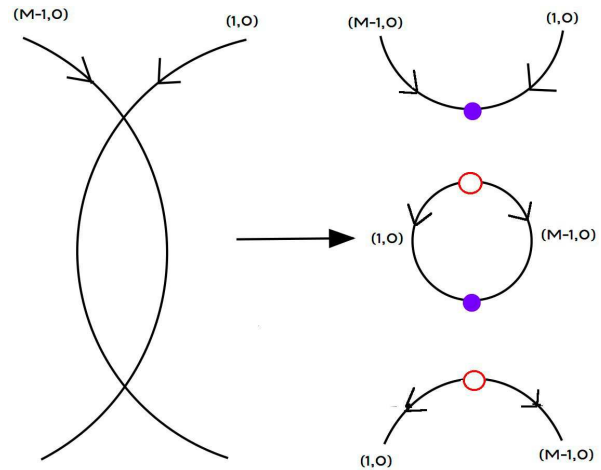}
\caption{This shows the intercommutation of a (M-1,0) string and a (1,0) 
string. 
The strings have the same tension but different charge. They form 
a Y-junction with an M-string with zero tension, which in some sense
invisibly connects the bead to the anti-bead. 
The beads are the finite energy objects that mediate the change of 
topological charge from one value to the other.  One can 
have beads and antibeads (conventionally chosen red/empty 
and blue/filled on the picture). We also show the formation of a 
closed loop via this process.}\label{m11}
\end{figure}

As in Ref. \cite{Jackson:2004zg}, when two strings of different types interact they form a Y-junction. Defining 
the quantum numbers $p$ and $q$ to be positive for a string that is directed 
outward from the junction, charge conservation implies that:
\begin{align}
\sum p &= n M\; , & \sum q & = 0\; ,
\end{align}
where $n$, the bead number, is valued in $\mathbb Z$ 
(negative $n$ implies an antibead). Thus, any 
interaction that results in $p \geq M$ implies that 
a bead - antibead pair must be formed to conserve charge (cf. Fig. \ref{m11}). In a fully general treatment, we should note, second order flux effects will come into play, complicating this story (and lifting the tension of the M-string above zero), though
we expect no qualitative changes to be wrought by these considerations \footnote{As stated, the tension formula for the (p,q)-string does not include
the contribution from the fluxes it sources.  This is usually a small, second order contribution,
 but in the case of the (M,q)-string -- where the first order effect vanishes
  --  it is the dominant one. This contribution slightly breaks the periodicity of the string tension.
This means that the M-string is actually a physical object and therefore while the 
process in Fig. 1 is still possible, one could also possibly form a junction with the very low tension ($T \sim (G\mu)^2$) M-string without forming beads. Given the tiny phase space for junctions yielding extremely low tension string states, we did not include this type of interactions in our analysis.
However, since this interaction will only further limit the production of beads, and thus 
make them even safer cosmologically, we do not consider this to be an important omission.
We also note that such interactions should not appreciably change the network dynamics.
Nonetheless, this deserves further study.}.

We can calculate the probability of bead formation by treating the 
formation of a bead-antibead pair as a tunneling event. Consider 
a simple triangular potential $V(x)$, with $V(0) = 2M_b$ and $V(L) = 0$ for
$L = 2M_b/\mu$. Using the WKB approximation, the probability 
of interaction is $P \sim e^{-\pi A}$, where A is the area under the curve. The factor of $\pi$ is included
to agree with a more precise calculation using an instanton method.  We then find that
\begin{align}\label{tunnel}
P_{\mbox{tunneling}} \sim e^{-\pi A}\sim e^{-2\pi \frac{M_b^2}{\mu}}.
\end{align} 
Interestingly, for the case of Fig. \ref{m11},  the probability is close to one for
an interesting range of $M$ values
\begin{align}
\pi A = \pi M_b^2/\mu_{1,0} =  M^2 \frac{2b^2g_s}{18 \pi \sin{\pi/M}}.
\end{align}
We see that $A\rightarrow \infty $ as $M\rightarrow \infty$, as we should expect, since
the bead mass grows faster than the string tension as a function of $M$. 
However, since the factor that multiplies $M^2$ may be of order $10^{-3}$ 
(if we assume $g_s \sim 10^{-1}$), we should have a non-suppressed
 production of beads for $M  \lesssim  10$.   Note that for smaller $g_s$ this minimum value will increase.
 
In Ref. \cite{Kachru:2002gs}, they argue that $M$ needs to be greater than $12$ in order
 for any anti-D3-brane to be classically stable in the KS throat.  Furthermore, we need $M$ to be greater than $\sim 1/g_s$ in order for the size of the extra dimensions to be big enough to trust the supergravity approximation. 
These considerations imply that bead production is always suppressed, albeit only slightly if $M$ is not much above
these bounds.

\begin{figure}[ht]
\centering
\psfrag{M-1}{$M-1$}
\psfrag{2}{$2$}
\psfrag{M-1}{$M-1$}
\psfrag{2}{$2$}
\psfrag{M-1}{$M-1$}
\psfrag{2}{$2$}
\psfrag{1}{$1$}
\includegraphics[width=8cm]{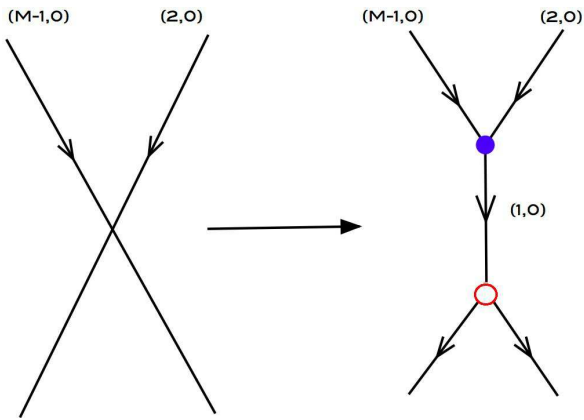}
\caption{An interaction between an (M-1,0)-string and a (2,0)-string 
forming a (1,0)-string. 
Note that the Y-junctions now have beads on top of them. Because of
that, it is no longer necessary for the tension to cancel exactly. 
If they do not the beads will accelerate.}\label{different}
\end{figure}

When two strings of different tensions meet and form beads (see Fig. \ref{different}),
 there are two different probabilities for tunneling since we have two different incoming string tension from which to choose. When this occurs in our calculations, we calculate the probability using the larger tension, since this will be the most probable outcome.  So we use the same formula as before, Eqn. \ref{tunnel}, with $\mu$ being the tension of the heavier of the two colliding strings.

Finally, we have the possibility of multiple beads getting stacked on each other through network dynamics. 
Though we expect this to be rare if beads are rare, it is easy to lay out how it would occur. 
Consider a bead living at a Y-junction. We generically expect  the bead to be moving, particularly 
since the growth of one of the connected strings is usually energetically favored. If this is the case, then
it would only be a matter of time before the bead at the junction would intersect any bead or anti-bead
that lives on the strings that are ``shrinking." When this collision occurs, the beads will interact, 
and we should expect their quantum numbers to add.

\subsection{Stability and D3-brane as Beads}

The F-string is charged under the NS-NS 2-form $B_2$. It has the standard Kalb-Ramond action
\begin{align}\label{kalb}
\mu_1 \int B_2\, ,
\end{align}
where $\mu_1 = 1/2\pi\ap$ is the F-string charge. The Kalb-Ramond action can be obtained from the action of a global Higgs string for distances much larger than the size of the core \cite{VilenkinShellard}. In 4D, the NS-NS 2-form is dual to an axion, so the Kalb-Ramond action just tells us that the F-string is emitting massless axions. This gives it the long-range interaction typical of global strings.

Global strings cannot break, so it is wrong to think of the F-string as ``breaking" on a bead.  What is happening, as we showed in Fig. (\ref{m11}) and (\ref{different}), is that two or more strings 
with different quantum numbers are \emph{joining} on a bead.  All strings still have long range interactions, so we can still detect their topological charge at infinity.  It is therefore impossible for the F-strings to disappear by ``breaking" on beads.  Note that even though each string individually has a boundary (where the bead is), the whole cosmic string (the cosmic necklace with the beads) does not.

Nevertheless something novel does happen at these junctions.  The strings that join at beads have different charges, and charge must be conserved at the intersection of the strings. A D3-brane wrapping a 3-cycle in the extra dimensions can do just that.  Indeed, as we mentioned before, the KS throat supports $M$ units of $F_3 = dC_2$ flux over a 3-cycle in the extra dimensions,
\begin{align}
\frac{1}{4\pi^2\ap^2}\int_{\mathcal{M}_3} F_3 = M.
\end{align}

The action (\ref{kalb}) is normally invariant under gauge transformation of $B_2$, $\delta B_2 = d\lambda_1$.  However, since we now have a boundary (for each part of the string in the necklace), this is no longer true.  The variation of the actions of all the strings ending on a given bead will add to 
\begin{align}
M\mu_1\int_t \lambda_1,
\end{align}
where the integration is over time.  This must be cancelled by the D3-brane action. The relevant term in the action is the kinetic term for the gauge field $F = dA$ living on the D3-brane, 
\begin{equation}
\frac{\mu_3}{g_s} \int_{M_3, t} \hf (B_2 + 2\pi\ap F) \wedge \star (B_2 + 2\pi\ap F)
\end{equation}
where $\mu_3 = \frac{1}{(2\pi)^3\ap^2}$ is the D3-brane charge. Varying this gives
\begin{equation}
- \frac{\mu_3}{g_s} \int_{M_3, t} \lambda_1 \wedge d \star (B_2 + 2\pi\ap F).
\end{equation}
By looking at the equation of motion for the gauge field, one finds that  $ \star (B_2 + 2\pi\ap F) = g_s C_2$, leading to 
\begin{align}
&= -\mu_3 \int_{M_3, t} \lambda_1 dC_2, \nonumber \\
&= - M \mu_1 \int_t \lambda_1.
\end{align}
This exactly cancels the contribution from the F-strings, showing that the beads are indeed wrapped D3-branes.

As global strings, one might expect the F-strings (as well as the D-strings) to be unstable 
to domain wall formation. The reason for this is simple: we have not yet seen a massless axion,
 and the presence of a massless scalar field is strongly constrained in cosmology (by, e.g., big-bang nucleosynthesis tests).  Therefore, the axion must pick up a mass.  We should expect to break the shift symmetry of the axion in doing so, creating a domain wall in the process.  The tension of the wall would then force the strings to shrink and disappear.

One can avoid this disastrous situation if the axion is ``swallowed" by a gauge field.  This can be accomplished in string theory with a Green-Schwarz-like coupling between $B_2$ and some gauge field \cite{Leblond:2004uc, Copeland:2003bj}.  The strings then become local instead of global,
 and the $B_2$ fields decay exponentially away from the strings, leaving no detectable long-range charge.  Polchinksi has conjectured \cite{Polchinski:2005bg} that,  in string theory, 
  all local strings of this type are unstable due to monopole formation. To see this in any particular case, we would have to look in detail at a complete string theory model where we could identify the monopole on which the string can break.  These monopoles would be different from the beads 
we discuss in this paper.  Unlike the domain wall instability, the monopole instability is a mild one:
it is quite possible for the monopole mass to be so high that the strings are long-lived and metastable.  For instance, the heterotic $SO(32)$ string, which exhibits this kind of breaking, is effectively stable for all practical, cosmological purposes \cite{Polchinski:2005bg}.  

Hence the question of stability is a model-dependent issue. For the remainder of this paper we 
will suppose that the strings are local and metastable, with a lifetime longer than the age of the universe. 
 Some progress has been made recently in identifying the monopoles that the strings would break
  on in the KS throat as a D5-brane configuration \cite{Verlinde:2006bc}. These monopoles have very large masses, in agreement with our working assumption of metastability. 
Furthermore, it shows that the beads we consider in this paper are not the same as the monopoles on which Polchinski's
conjecture predicts the string would be unstable to breakage. These developments suggest that
there is much more work to be done in understanding the underlying
 topological structure of these strings.

\subsection{Binding}

In general, the tension formula for strings in a warped throat (Eq. \ref{finalanswer}) leads to 
higher binding energies than in flat space. Ongoing numerical work \cite{Wyman:2006a} has suggested
that vortex binding may occur efficiently in a gauge theory model. 
 A recent analytical study of the kinematics
of string binding \cite{Copeland:2006etd} confirms that bound states can indeed
grow relativistically. We thus retain the working
 assumption from Ref. \cite{Tye:2005fn} that strings bind instantaneously.
The new structure in the $(p,q)$ tension spectrum (Eq. \ref{finalanswer})
leads to a different set of binding rules than is described in Ref. \cite{Tye:2005fn}.
The most important difference is that non-coprime $(p,q)$ states are now 
bound (these are states for which $p$ and $q$ have a common factor; 
i.e., $p = kN, \, q=lN$).  In that paper, the fact that these states were not
bound was important in preventing UV-catastrophes in the network evolution (For explanation, see
 note at end of bibliography,
\footnote{By UV-catastrophes, we mean the fact that there is no upper bound to string tension. Thus, 
we can imagine cases where strings continually bind together, increasing the quantum number with each interaction, 
causing the tension to evolve to ever-higher values. This sort of ever-upward evolution would be 
cosmologically disastrous, as very-high-tension strings would come to dominate the universe's energy density.} ).
A related change is that $(p,0)$ states with $p>1$ are also now permitted to
exist.  $(0,q)$ states are not bound, however, and cannot form.
Adding all these new states and removing an important string binding rule (the decay
of non-coprime states) reopens the problem of cosmological UV catastrophes.
We tested our model with these assumptions and found that the
cosmological density did continue to grow to late times. However, perhaps these new
rules are too simple. Non-coprime string states still have much less binding energy than coprime
states when $p\ll M$. To account for this, we prevent some fraction of non-coprime 
states from forming by allowing them to ``decay" immediately upon formation as before. Let us define
\bas
\mathcal{E}^{\mbox{unbound}}_{(Nk,Nl)} & = & N \mu_{(k,l)},\\
\mathcal{E}^{\mbox{bound}}_{(Nk,Nl)} & = & \mu_{(Nk,Nl)}, \\
P^{\mbox{decay}}_{(Nk,Nl)} & = & \exp \left( {\mathcal{E}^{\mbox{bound}}_{(Nk,Nl)}  -\mathcal{E}^{\mbox{unbound}}_{(Nk,Nl)}} \over {\mathcal{E}^{\mbox{bound}}_{(Nk,Nl)} } \right ). 
\eas
We use $P^{\mbox{decay}}$ as a branching ratio between the formation of non-coprime
states and their coprime daughter states. This immediately restores network scaling. 

\section{Network Evolution}\label{network}
To study the evolution of string networks in a curved background, we modified the 
network model described in Ref. \cite{Tye:2005fn} to include the new features described 
above. Our review of this model will be terse; readers looking for more detail
are referred to the original paper, where it is discussed at far greater length. 
The new features we have added to that model are as follows.
\begin{itemize}
\item We have introduced a maximum value, $M$, 
for the quantum number $p$. Thus, we take $(k+M,l) = (k,l)$,
so that binding interactions that seem to form a $p = k+M$ state actually populate the state 
with $p=k$. 
\item When we attempt to form a $p = k +M$ state, we must take into account 
bead formation. This introduces a tunneling suppression factor related to the bead mass; see Eqn. \ref{tunnel}.
\item The set of allowed states now includes non-coprime $(p,q)$ states 
as well as all $(p,0)$ states, since they are now bound. 
\item Non-coprime states are allowed to decay, upon formation, with a probability $P^{\mbox{decay}}$.
\item  The new tension formula, Eqn. \ref{finalanswer}, is used in the calculation
of interaction probabilities. 
\item We neglect bead dynamics in the evolution of the network.  The justifications for this are 
detailed in the next section. 
\end{itemize}

\subsection{The MTVOS model}
Following Ref. \cite{Tye:2005fn}, we define an average
string velocity, $v$, and single network string correlation length-scale, $L$. 
Each are shared by all strings in our network, regardless of their tension or number density.
Using the cosmological scale factor, $a$,
we can also introduce a conformal time by $d\eta=vdt/a$ and a comoving string length, $\ell=L/a$.   Denoting each string $(p,q)$ state by a single Greek letter for compactness, 
we can further define a scaled string number density,
$$
\Nalpha = a^2 \nalpha = \frac{a^2\rho_{\alpha} \sqrt{1-v^2}}{\mu_{\alpha}},
$$
and a string interaction probability $P_{\alpha\beta\gamma}$ (the probability that 
the interaction of a string of type $\beta$ with a string of type $\gamma$ will
produce a string of type $\alpha$). Since it is a key point, we restate it for emphasis: 
in this model, the correlation length $L$ 
is shared by all $(p,q)$ states. It is not related to the physical distance between strings
(The distance between two strings of type $\alpha$ may be given roughly in terms
of that state's number density; $\sim \Nalpha^{-1/2}$) .
We may then assemble a set of network evolution equations for evolving each of
these string states, the overall correlation length, and the average string velocity.
To connect conformal time to physical time, we 
assume a Friedmann evolution for $a$ in a radiation or
matter-dominated universe. We take
 into account string loop formation (the terms proportional to 
$c_2$ and $P$); string binding (proportional to $F$); string stretching ($c_1 \eta$); 
Hubble friction for reducing $v$ ($\propto H a$); and acceleration 
of $v$ due to string straightening ($\propto \ell^{-1}$).
\ba
\Nalpha' & = &-{c_2\Nalpha\over\ell}-{P\Nalpha^2\ell}
+F\ell\left[{1\over 2}\sum_{\beta,\gamma}
P_{\alpha\beta\gamma}\Nbeta\Ngamma
(1+\delta_{\beta\gamma})\right. \nonumber\\
&& \left.-\sum_{\beta,\gamma}
P_{\beta\gamma\alpha}\Ngamma\Nalpha(1+\delta_{\gamma\alpha})\right]~, \\
\ell & = & \ell(0)+c_1\eta ~,\\
v^\prime& = &{(1-v^2)\over v}\left(-2Hav+{c_2\over\ell}
\right)~,
\ea
with primes denoting derivatives with respect to conformal time;
the constants $c_1 = 0.21$ and $c_2=0.18$ are obtained
by matching the scaling-era values of $HL$ and $v$ to results from old string network
 simulations, while the constants $F$ and $P$
are generally left as variable parameters. The constant $F$ indicates the overall strength
of inter-string binding interactions; we take $F=1.0$ for all our integrations. 
The constant $P$ indicates the string intercommutation and loop production
rate.  Note that $P$
is not directly related to the probability of string intercommutation. It is better
thought of as a loop formation efficiency parameter. 
Based on the results in Ref. \cite{Tye:2005fn}, we know that for
non-binding networks (where $F \rightarrow 0$), we must take $P=0.28$
to match to simulations. Hence, we simply assume this value for $P$ in all of the
present calculations. Varying $P$ changes the overall number of strings
present in the simulation -- see \cite{Tye:2005fn} for discussion --
but affects the results for beads minimally.

\subsection{New Binding Physics}
All of the new physics introduced by the new
warped geometry under consideration are encoded in the string interaction probability
term,  $P_{\alpha\beta\gamma}$. First, we write down the general form of the formula for this,
 given in Ref  \cite{Tye:2005fn}, taking $\beta = (k,l)$ and $\gamma = (m,n)$ to be the two
 interacting string states:
\be
P_{\alpha \beta \gamma}^{\pm} = \frac{1}{2} \left ( 1 \mp \cos \theta^{crit}_{klmn} \right ),
\label{pbasic}
\ee
where $P^+$ indicates $\alpha = (p,q) = (k+m,l+n)$ and $P^-$
indicates $\alpha = (p,q) = (k-m,l-n)$. The critical angle that differentiates between the two type of binding is:
\begin{align}
\cos \theta^{crit}_{klmn} &= {e_{kl} \cdot e_{mn} \over |e_{kl}| |e_{mn}|}, \nonumber\\
e_{mn} &=(\frac{bM}{\pi} \sin(\frac{\pi m,}{M})g_s, n).
\end{align}
The use of this critical angle encodes the physics of the new tension formula. 
The next things to include are the suppression of string interactions that form beads
and the periodicity of the string tension states.
We include these effects in $P$ in the following way:
$$
P_{(p+M,q) (k,l) (m,n)} =  P_{\mbox{tunneling}} P_{(p,q) (k,l) (m,n)},
$$
where $ P_{\mbox{tunneling}} = e^{-\pi M_b^2/\mu_{max}}$ is only included
when the bound state being formed has quantum number $p \geq M$. 

Finally, we include the probability of decay for non-coprime states with small binding energy.  Recall that
this is necessary because in a warped geometry non-coprime states
are slightly bound. The decay of non-coprime states was very important for the regulation
of UV catastrophes in previous integrations of these equations \cite{Tye:2005fn}.  We originally
thought  that the periodicity of $p$ might help to regulate these catastrophes; 
early numerical experiments told us that this alone was not enough. To implement this decay, we take
$$
(Nk,Nl) \rightarrow \left \{ \begin{array}{cc}
N(k,l) & \mbox{with probability } P^{\mbox{decay}} \\
(Nk,Nl) & \mbox{with probability } 1 - P^{\mbox{decay}} 
\end{array} \right .
$$
In practice, we modify our network interaction term so that, when $\alpha=(Nk,Nl)$,
$$
P_{\alpha\beta\gamma}  \rightarrow  (1-P^{\mbox{decay}}_{(Nk,Nl)})P_{\alpha\beta\gamma} + 
 P^{\mbox{decay}}_{(Nk,Nl)} P_{N(k,l) \beta \gamma}\; .
$$
Note that the overall probability
does not add up to one anymore.  Effectively, the factors $P_{\mbox{tunneling}}$
and $P^{\mbox{decay}}$
reduce the overall 
factor of F for certain string interactions. Since these reductions only apply for some specific
interactions, it is more convenient to include these effects in
 $P_{\alpha \beta \gamma}$ than to make F also a function of the
 string quantum numbers.
 
 \subsection{Network Results}

As we alluded to before, the immediate result of our early network integrations was
 that some sort of decay of the slightly-bound states is necessary 
 for the networks to go to scaling. Without
any decay mechanism, the number of allowed states appears to be too large,  
leading to a UV-catastrophe, albeit a mild one. The probabilistic 
decay of non-coprime states, however, very easily regulates these catastrophes. 

\begin{figure}[htbp] 
   \centering
   \includegraphics[width= 8cm, height=10cm]{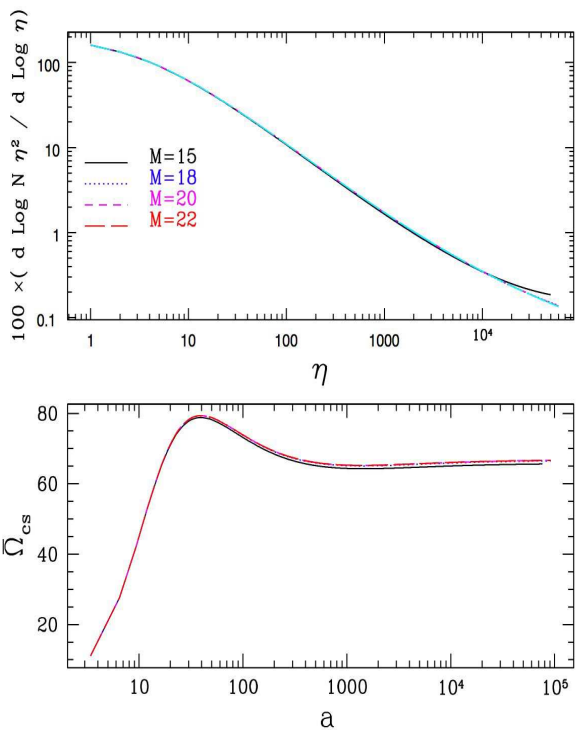} 
   \caption{The bottom panel shows the evolution of 
$\tilde{\Omega}_{cs} = \Omega_{cs} / ((8/3) \pi G \mu_{(0,1)})$ 
for various values of M, taking the string coupling $g_s=0.5$. The top panel shows the rate of change in the comoving number density $N \eta^2$; in the scaling regime, 
$d \log N \eta^2 / d \log \eta = 0$. }
   \label{omegas}
\end{figure}

\begin{figure}[htbp] 
   \centering
   \includegraphics[width=8cm]{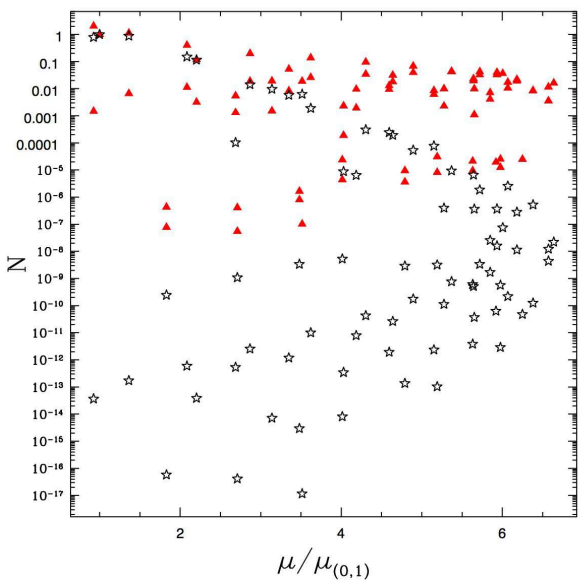} 
   \caption{The final scaling-era spectra for $M=20$ taking the string
   coupling $g_s=1.0$, with $N_{(0,1)}$ normalized to unity and the other number densities
  scaled accordingly. The black stars are for the standard initial conditions; the
  red triangles are for initial conditions in which the $p>M/2$ states are initially
  populated. String spectra for other values of M are similar.}
   \label{mus}
\end{figure}

\begin{figure}[htbp] 
   \centering
   \includegraphics[width=8cm,height= 9cm]{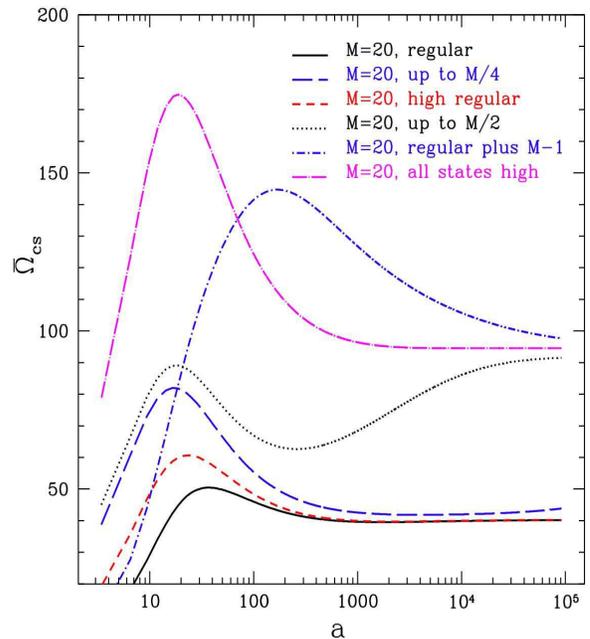} 
   \caption{The evolution of 
$\tilde{\Omega}_{cs} = \Omega_{cs} / ((8/3) \pi G \mu_{(0,1)})$ 
for various initial conditions, taking the string coupling $g_s=1.0$ and M = 20. The black, solid
line: only (0,1) and (1,0) states are initially populated. The blue,
long-dashed line: all states with $p<M/4$ initially populated.
The red, dashed line:  only (0,1) and (1,0) states, but with the initial populations
 quadrupled in size. The black, dash-dotted line:  all states with $p\leq M/2$ populated. The blue, dot-dashed line: only the (0,1), (1,0), and (M-1,0) states are initially
populated. The magenta long-dashed/dotted line: 
every allowed $(p,q)$ state is begun with a high
initial population.We see that there are two scaling values of $\tilde\Omega$: 
$\simeq 40$, for cases where the initial conditions have only the low values of $p < M/2$ initially populated; and $\simeq  95$ when the states with $p \geq M/2$ are initially populated.}
   \label{diffics}
\end{figure}

The chief results of this network model are quite similar to those of
the simpler model studied in \cite{Tye:2005fn}. We should note that another group has proposed
a related but more detailed model, which also returns very similar results \cite{Avgoustidis:2007a}.  In the numerical integrations
presented here we allowed $p$ to range between $0$ and $M-1$ and $q$ between
$0$ and $3$; we also always assume $F=1$ and $P=0.28$. Varying the maximum value of $q$ had little effect on the results, 
while leaving out states with $p \lesssim M$ can lead to serious errors. Our results:
\begin{itemize}
\item Networks are still able to scale.
\item Networks reach scaling quite quickly for each value of $M$ that we tested (see top panel of Fig. \ref{omegas}). Larger values of $M$ could not be calculated due to the large number of states involved. However, we expect to recover the results of Ref. \cite{Tye:2005fn} for networks with very large values of $M$. Since we would expect any novel network behavior to be most manifest at low values of $M$, it was sufficient to test just a few such values. The dependence on initial conditions, discussed below, is
an instance of such a behavior.
\item For a given set of initial conditions, the overall density in cosmic strings does not depend on M. For the simplest initial conditions (equal numbers of (0,1) and (1,0) strings being the only states 
initially populated) we find, for $g_s=1$,  $\tilde{\Omega}_{strings} \simeq 40$ for all runs (see Fig. \ref{omegas}); for $g_s=0.5$, we find  $\tilde{\Omega}_{strings} \simeq 70$, where we have 
defined $\tilde{\Omega}_{strings} \equiv \Omega_{strings} / ((8/3) \pi G \mu_{(0,1)})$
\item String density is still very dependent on string tension: we again find  
$N_{\alpha} \propto \mu_{\alpha}^{-n}$, and $n \sim 8$ for standard initial conditions.
 However, this relation is now more an envelope function than 
a predictive function: the sinusoidal shape of the $p$ tension leads to greatly-enhanced
structure in the string spectrum (see Fig. \ref{mus}). For instance, for M=15, the tension 
of the (1,0) strings and the (14,0) strings are identical; but, for our standard initial
conditions, $N_{(1,0)} / N_{(14,0)} = 2.5\times10^6$.
 See table \ref{tab:ratios} for more such comparisons. 
\item The complexity of the string tension spectrum leads to a dependance on initial
conditions in $\tilde{\Omega}_{strings}$. Essentially, the sinusoidal shape of the tension formula
for $p$ strings suppresses the formation of $p>M/2$ states if the initial distribution 
of string states is confined to $p<M/2$. However, if the $p>M/2$ states have similar initial populations
to the $p<M/2$ states, then the scaling value of $\tilde{\Omega}$ is approximately doubled. For instance,
for $g_s =1$, the scaling value of $\tilde{\Omega}$ jumps to 95 when states with $p>M/2$ are initialized.
See Fig. \ref{diffics} for an example of this effect.
\item The tension spectrum is also bimodal in initial-condition response. 	The results 
pictured in Fig. \ref{mus} show the tension spectrum for initial conditions where the $p>M/2$ states
are initially populated. In that case, the envelope-function for number density versus tension is
much less steep: $n \sim 3$ or $4$. 
\end{itemize}

\begin{table}
\begin{center}
\begin{tabular}{|c|c|c|c|c|}
\cline{2-5}
\multicolumn{1}{c}{} \vline& \multicolumn{2}{c}{$g_s = 0.5$} \vline &  \multicolumn{2}{c} {$g_s = 1.0$} \vline \\
\cline{2-5}
\hline
M & ${N_{(1,0)} \over {N_{(M-1,0)}}}$ &$\frac{N_{(1,\pm1)}}{N_{(M-1,\pm1)}}$  & $\frac{N_{(1,0)}}{N_{(M-1,1)}}$ & $\frac{N_{(1,\pm1)}}{N_{(M-1,\pm1)}}$ \\
\hline
15& $2.5\times10^6$ & $1.6\times10^5$  & $1.8\times10^{11}$ & $3.4\times10^{8}$\\
18 & $5.5\times10^7$ & $1.2\times10^7$ & $3.8\times10^{11}$ & $9.4\times10^{10}$\\
20 & $1.1\times10^9$ & $2.2\times10^8$ & $2.2\times10^{13}$ & $5.1\times10^{12}$ \\
22 & $2.1\times10^{10}$ & $3.9\times10^9$ & $1.2\times10^{15}$ & $2.7\times10^{14} $\\
15, $p>M/2$ & $1.1\times10^{3}$ & 35 & $4.5\times10^{4}$ & $100$\\
20, $p>M/2$ & 624 & 54 & $1.4\times10^{3}$ & $170$\\
\hline
\end{tabular}
\end{center}
\caption{\label{tab:ratios}The relative populations of the lowest-tension $p=1$ and $p=M-1$
states. The first four rows were calculated under the standard initial conditions, for which only $p<M/2$ states are initially populated. The last two are calculated under initial conditions where all
states, including $p>M/2$, are initially populated.
Note that $\mu_{(1,q)} = \mu_{(M-1,q)}$. In all cases, the $(M-1,\pm1)$ state had a larger
population than the $(M-1,0)$ state.}
\end{table}

\subsection{Other Considerations}
We also considered
including an interaction term that took into account the strings' kinetic energy in determining 
whether a particular binding interaction would occur. We wrote down a suppression term
similar to our expression for $P^{\mbox{decay}}$, except with a denominator based
on the interacting strings' kinetic energy. Evolving networks with this suppression
term did regulate the most obvious uncontrolled growth of cosmic string network density,
even when we did not allow the decay of non-coprime states.
However, the networks did not enter a scaling regime: instead, they continued to evolve to
late times, not becoming dominant but with an always slowly increasing 
cosmological density. The reason for this
is that kinetically suppressing string interactions slightly favors the growth of
 high-tension number density.  Heavy strings have a great deal of kinetic energy. If strings
 with high kinetic energy cannot easily undergo binding interaction,
 heavy strings become isolated; their  effective interaction probability decreases, $F \rightarrow 0$. Since phase space and binding energy effects
both tend to decrease high-tension-state number density, we get a competition. 
For values of $M\lesssim15$, this competition leads to late-time network evolution, 
even when the decay of non-coprime states is allowed to proceed.
The MTVOS model, as formulated, is difficult to evolve to late times, since steps
in conformal time $d \eta \propto a^{-1}$. It may be worth reformulating the equations
to study this late-time network development, which may provide additional string network
phenomenology that would be interesting to study further. For the time being, though, we have simply
omitted any kinetic energy-based suppression from our model.

While this paper was being prepared,
 Copeland, Kibble, and Steer \cite{Copeland:2006etd} published
a detailed analysis of kinematical constraints on string binding, building
on some results which they had previously published \cite{Copeland:2006eh}. Their formulae
for kinematical determination of whether binding proceeds are derived
in the Nambu-Goto string limit for several classes of string binding configurations. 
Since our network model needs binding probability formulae that cover all possible
binding events, we were not able to apply their formulae to our model. However,
two of their chief results -- that lighter strings are preferred by kinematics and
that string binding, when it occurs, proceeds relativistically -- are in line with the
assumptions and conclusions of our work. Any future development 
of the MTVOS network equations should take into account the constraints
that Copeland et al. have derived, including their finding that
binding networks may exhibit smaller r.m.s. string velocities. Ultimately, a fusion of their approach
with the approach taken here could provide a robust framework
for the quasi-analytical study of binding string networks.
\vspace{-14 pt }
\subsection{Populating Low-Tension, High-$p$ States}

In Ref. \cite{Tye:2005fn}, it was proposed that the main observable difference between a 
network of (p,q)-strings and a single string network is a factor of 3 in $\Omega_{cs}$.  The reasoning
behind this is that the network reaches scaling with significant populations in only its lowest-lying
tension states -- those not liable to decay.  For the $(p,q)$-strings network in flat space, these are the $(1,0)$, $(0,1)$ and $(1,1)$ states -- a total of 3 states leading
to a factor of 3 in energy density as compared to a single-species string network.  
For a general network with ``m" types of strings, the density of strings should, then, 
 be $2^m -1$ times the results for a single string network.

In our case we have two additional low tension states, $(M-1,0)$ and $(M-1,1)$. This, too, leads to an
enhanced density of strings. However, this enhancement is appreciable only if these states
are populated initially, rather than being formed only through dynamics. When these sorts of initial 
conditions are considered, we can see from Fig. \ref{diffics} that the overall network 
density is approximately doubled, just as this rule of thumb predicts.

Initially populating high quantum number states is, however, probably unrealistic. 
Also, these initial conditions may already be excluded, since highly populated $(M-1,n)$ states can lead 
to problems with overproduction of beads. We will discuss this concern more below.

\section{Beads on a String} \label{beadanalysis}

As we have argued before, whenever two strings meet such that the absolute value of the sum of their $p$ quantum
numbers exceeds $M$, a bead must be formed for the interaction to proceed.
The beads that are formed do not exist on their own. They are constrained to live on
their confined flux tubes -- to be threaded by the string network on which they were formed.
Since the $0$-string and the $M$-string have zero tension,
the beads are never a monopole-like ``end" of a string. Instead, they must 
function as  ``beads" on the string, always connecting two different kinds of string \cite{Hindmarsh:1985xc}.
As mentioned previously, bead production is a tunneling process, and therefore it can be totally
suppressed if the bead mass is very large. When $M$ is not too large, though bead
production cannot be neglected.  We might worry that beads, as point-like
sources of energy-density, could cause serious problems for the cosmology of this model. 
Indeed we know that ordinary monopoles can overclose the universe, since their densities 
decrease only like $1/a^3$ and hence grow rapidly with respect to the radiation 
component of the universe until they are dominant.
 In our case, since these monopoles -- or beads -- are stuck on strings, we expect a very different behavior. 
A useful parameter to study this is $r = \frac{M_b}{\mu d}$,
where $d$ is the average distance between beads \cite{Berezinsky:1997td}.

For this first analysis we shall consider only a single cosmic string tension. 
The average mass per unit length of such a necklace is $(r+1) \mu$.
For $r \ll 1$, the beads contribute a negligible amount to the mass density on a string
and we can suppose that the beads are simply carried around and remain cosmologically 
innocuous. For $r \gg 1$, 
the beads are the dominant contribution to the mass density and the network 
should look like a  gas of monopoles. This is the cosmologically disastrous case.
Intermediate values of $r$ ($r \sim \mathcal{O}(1)$) are interesting, because they may not be ruled 
out, yet would imply a higher effective string tension and energy density for the string network.

In the cosmological model presented in this paper we start with a network of $(p,q)$
strings.  These strings produce beads only through interactions -- there are no primordial beads.
Hence we necessarily start with $r \ll 1$. What we need to calculate is how this dimensionless number evolves with the development of the network. 
A simple model due to Berezinsky and Vilenkin describes the salient features \cite{Berezinsky:1997td}.

\subsection{Evolution in the Scaling Regime}

To begin simply, we will assume that a cosmic supernecklace network has reached
a scaling solution without having formed too many beads, so that it enters the
scaling era with $r\ll1$.
We can then study the time evolution of $r$ by looking at energy
conservation for a particular (closed) necklace of length $L = Nd = \frac{N M_b}{r\mu}$,
where N is the total number of beads on the necklace, which we take to be a constant.
The size of this loop-necklace can be greater than the Hubble length; if so, 
the closed necklace will appear to be an infinitely long string in any particular Hubble volume. 
In the spirit of Ref.  \cite{Berezinsky:1997td}, we can then write an energy balance equation for this closed necklace:
\begin{align}\label{firstlaw}
\dot E = -P \dot V - \dot E_g,
\end{align}
where $E$ is the total energy of the necklace, $P$ is the effective pressure 
and $\dot E_g$ is the energy loss by gravitational radiation from small scale wiggles. This approach neglects several things.
\begin{itemize}
\item The production and annihilation of beads. We will return to this point later and discuss
the effect of including bead production.
Annihilation could be an important effect when $r$ approaches 1 while unimportant
in the scaling regime which we discuss below, where the inter-bead distance is of order the Hubble scale. However, it might be important in the transient regime. Given the uncertainties, 
we did not include annihilation in our analysis. Inclusion of annihilation will only aid in preventing 
bead domination of the network's dynamics.
\item Energy lost to gravitational radiation from moving beads.  Assuming $r\ll 1$ and that no average force acts on the beads should keep this contribution negligible.
\item Loop formation. We will come back and include the effect of loop formation later.  Loop formation 
can be very important. In fact, we will argue that it leads to a scaling value for $r$; but to get to that point we 
need to understand the evolution of $r$ in the absence of loops first. 
\item  The physics of Y-junctions, which can ``capture" beads and transport them around as we have discussed briefly
and will touch on again below.  We assume this will not change the physics too much, but we have
not yet been able to quantify this assumption.
\item Finally, the fact that the work term $-P\dot V$ is really only thermodynamically well-defined
for an ensemble of cosmic necklaces. Nevertheless, we can
connect this averaged quantity to a particular
loop by considering the work done per cosmic necklace in the scaling regime. 
This is, of course, a rough approximation.

\end{itemize}
Eq. (\ref{firstlaw}) contains two terms through which the energy of a necklace can change.
The first term is the work done by the Hubble expansion. Depending on the sign of the pressure,
this can cause either a gain or a loss of necklace energy.  The second term is the loss of energy due to gravitational radiation.  Below, we will assume that most of the gravitational radiation is coming from the small wiggles on the string.

\subsubsection{Gravitational Radiation}

The power emitted by the wiggles on a string is approximately given by \cite{VilenkinShellard}
\ba 
\frac{dP}{dl} \sim 2\pi G(\tilde\mu -\mu)^2\bar\omega
\ea
where $\bar\omega$ is the average frequency of the wiggle and $\tilde\mu$ is the effective tension of the string.
Estimates of the effective tension of a wiggly string vary, from $2\mu$ in standard lore \cite{VilenkinShellard}
to $1.1\mu$ in more recent analytical calculations \cite{Polchinski:2006ee}.  The precise value
 turns out to have very little effect on our result. Nevertheless, 
 we leave it as a free parameter and write $\tilde\mu = \beta\mu$. 
We take the average wavelength of the wiggles to set the size of the loops $l_{\text{loop}} = \alpha t$ 
and therefore $\bar\omega = 2\pi/(\alpha t)$. With all this we find
\ba
\frac{dP}{dl} & \sim &\frac{4\pi^2 G\mu^2(\beta -1)^2}{\alpha t}\; ,\nonumber\\
P & = &\dot E_g = \frac{4\pi^2 G\mu^2 (\beta -1)^2 L}{\alpha t}\; ,\nonumber\\
& = &\kappa_g \frac{\mu Nd}{t} = \kappa_g \frac{N M_b}{r t}\; ,
\ea 
where $L = Nd$ and  $\kappa_g = \frac{4\pi^2 G\mu(\beta -1)^2}{\alpha}$ .

\subsubsection{Hubble Stretching}
To calculate the Hubble stretching of a single cosmic necklace, we first calculate the Hubble work 
on the network as a whole during scaling. Using this result, we can approximate
\ba
(-P\dot V)_{\text{1-necklace}} \approx (-P\dot V)_{\text{total}}\frac{E_{\text{1-necklace}}}{E_{\text{total}}}\; .
\ea

Assuming that $r \ll 1$, we can suppose that the pressure is mostly due to a gas of cosmic strings.
We then have
\ba
P &=& \frac13 \rho_s (2v^2 -1)\;,
\ea
where $\rho_s \sim \frac{(r+1)\mu}{\xi^2}$ during scaling and $\xi$ is the correlation length in the scaling regime. 
For a network of cosmic necklaces, we expect  $\xi \sim (r+1)^{\hf} t$ \cite{Berezinsky:1997td}. 
Taking the total volume to be a comoving volume characterized by a length scale, L, we get 
\ba
(-P\dot V)_{total} = \frac{\mu V \kappa_s}{\xi^2t}\;,
\ea
 where $V/\xi^2$ is the total length of strings in the comoving volume V, $\kappa_s = (1- 2v^2) \nu$ and we took $a \sim t^\nu$.  The total energy of comic necklaces in a comoving volume V is $E = (r+1)\mu V/\xi^2$ and the energy of the specific cosmic necklace we are looking at is $E = (r+1)\mu L$.  Hence the Hubble work for 1-necklace of length L is
 \ba
 (-P\dot V)_{1-necklace} = \frac{N M_b \kappa_s}{r t} \approx \frac{N M_b \kappa_s}{rt}\; .
 \ea
This is of course an approximation, but it has the correct asymptotic behavior, i.e., it grows for smaller value of r.

Plugging everything in the energy balance equation (Eqn. \ref{firstlaw}), we obtain 
\ba
\frac{\dot r }{r} = -\frac{\kappa_s}{t} + \frac{\kappa_g}{t}.
\ea
Using the scaling value for $v^2$ from simulations \cite{Bennett:1987vf, Allen:1990tv},
we find that $\kappa_s$ is $0.07$ and $0.14$, in the radiation and matter eras, respectively. This gives a negative cosmic string pressure, so the strings are stretched by the Hubble expansion.  Thus, the energy balance equation for the necklaces 
expresses a competition between Hubble stretching and shortening due to gravitational radiation.  
Note that at $r=0$ and/or $t=0$, the preceding equation fails.  This is not a serious defect, 
just a limitation caused by our way of formulating the problem. This formulation is convenient for us, and
since we will always consider $r>0$ and can take our initial time (time at the end of inflation) to be non-zero,
it will not cause any problems for our analysis.

\subsubsection{Solution and Interpretation}
\label{revolution}

The solution to this equation is
 $r(t) \sim t^{\kappa_g-\kappa_s}$. This result can be understood as follows. 
 The stretching of the string ($\kappa_s$) tends to decrease $r$, whereas
the radiation from the small scale structure ($\kappa_g$) tends to shorten the string
and increase $r$.  

In Ref. \cite{Berezinsky:1997td}, they assumed that $\kappa_g \sim 1$, which follows from the
assumption that $\alpha \sim G\mu$. This predicts an $r$ that grows until it becomes a value of order $1$,
at which time the various approximations done in this calculation break down. 
If, on the other hand, we take at face value the recent results from \cite{Vanchurin:2005pa, Olum:2006ix}
that give $\alpha \sim 0.1$, we then find $\kappa_g \ll 1$ and $r$ decreases with time. 
We should also note that other recent expanding space simulations similarly find considerable
loop production at length-scales much larger than the gravitational back-reaction
scale \cite{Martins:2005fp, Ringeval:2005ce}.
In this scenario, as $\alpha$ grows, the small-scale structure gets bigger, and we thus get
less gravitational radiation for a given length of string.  The stretching then wins over the 
shortening and we find that the beads will get farther and farther apart.

This conclusion is also consistent with the recent analytical results of Ref. \cite{Polchinski:2006ee},
who found that stretching is more important than what was previously anticipated from simulations.
Given these successes, we claim that this approach, though clearly very simplified,
captures the essence of the problem. On a given string we should expect the beads 
to get farther and farther apart if the string stretching from Hubble expansion 
is more important than the string shortening from gravitational radiation.  Of course, 
the value of $\alpha$ is still a subject of debate. It is hard completely to determine the status of the 
problem without a more in-depth analysis.  A reader wishing to pursue these questions more
will find a good starting point for such a detailed analysis in Ref. \cite{Siemens:2000ty}. 

We can also take an alternative approach and try to determine a range of $\alpha$ values
 such that the stretching dominates over the shortening. From the numbers given above, this range
 is given by
\ba
\alpha \gtrsim 10^3 \, G\mu (\beta-1)^2\; .
\ea
With $G\mu \sim 10^{-7}$ this gives roughly 
\begin{eqnarray}
\alpha & \gtrsim  &10^{-4}\;\;\;\; \mbox{for } \beta = 2, \mbox{or}\nonumber \\
\alpha & \gtrsim  & 10^{-6}\;\;\;\; \mbox{for } \beta = 1.1
\end{eqnarray}
These values for $\alpha$ lie in the ``middle ground" between the various claims in the literature,
so finding the real value of $\alpha$ is critical for determining the fate of beads on cosmic necklaces.

\subsubsection{Including Production}

One can include the dynamical production of beads in the previous analysis
 by allowing N to be a function of time.  Since we want to find conditions for which 
 beads are subdominant, we take $r\ll 1$ as an ansatz. Using this, we find:
\ba\label{fulleqt}
\frac{\dot r }{r} = -\frac{\kappa_s}{t} + \frac{\kappa_g}{t}+r \frac{\dot N}{N}\; .
\ea

When we reach the scaling regime,  a certain number of beads have been formed. The network has settled into a stable state. The great majority of the strings are in the lowest energy states: $(1,0)$, $(0,1)$ and $(1,1)$ (and, possibly, $(M-1,0)$ and $(M-1,1)$, depending on initial conditions).  In the
former case, it is only rare interactions that produce beads.  We can estimate this rate in the scaling regime very simply. The number of collision per unit of time is $\sim 1/t$ (for strings going at speed $v\sim 1$ and separated by $L\sim t$), and the number of beads formed per unit of time is the number of such collisions that form beads.  Looking at Table \ref{tab:ratios}, we can guess that the
dominant bead-forming interactions will be between $(1,n)$ and $(M-1,n)$ strings, where $n=0,1$.
The smaller population of the $(M-1,0)$ states will be the first source 
of bead production suppression. As a proxy for this effect, let us define,
$$
R \equiv \frac{N_{(M-1,0)}+N_{(M-1,1)}}{N_{(1,0)} + N_{(1,1)}},
$$
 (note that $R$ is related to the inverse of the ratios as given in Table \ref{tab:ratios}).
 The second source of suppression is the tunneling suppression. Together, these imply a rate
of production given by something like
\ba
\dot N & \sim & \frac{R}{t}  e^{-P_{\text{tunneling}}}\; , \\
& \sim & \frac{R}{t}  e^{-M^2}.
\ea
There is no clear relationship between $M$ and $R$, but for most scenarios we should expect
$R$ to decrease for increasing $M$; it will certainly never increase drastically.
Thus, tunneling suppression forces this rate to zero for 
large values of $M$.
The rate also decreases at late times, when strings gets farther and farther away from each other and collisions become rarer. 

Still, $N(t) \sim \ln t$, so the number of beads always grows -- just very slowly. 
Thus, we can make the term $\dot N /N$ negligible in (\ref{fulleqt}) because of its very small coefficient. 

Although production is negligible in the scaling regime, it will be important during
the transient regime. One might reasonably fear that the presence of too 
many large quantum number strings at the onset of the network evolution 
might create too many beads.

\subsection{Loops and the Scaling of $r$}
\label{loops}

In this section we discuss the effect of loop formation in the scaling regime.
Loops can have a very important effect on the value of $r$. In fact, loop formation should promote a scaling solution in some cases. To see this, we first observe that each
loop that is formed removes an amount $\ell$ of string length. When a loop is
formed, there may or may not be beads that are trapped on the loop. If there are no
beads on the loop, then the loop formation will decrease the distance between 
beads, increasing $r$.  If there is a bead and a anti-bead on the loop-forming portion of the string and the average interbead distance is half the average size of loops, then $r$ will be unaffected. 

If $r$ would decrease from stretching in the absence of loops, adding 
loop formation can either slow down the decrease of $r$ or cause it to cease decreasing altogether.
  The latter is the interesting case.
  
In necklaces with sparse beads, loop formation removes length between beads.
As this process goes on, the average spacing between beads
will eventually become of order half the average loop size, $\ell = \alpha t$. At this time,
 we will expect nearly every loop to contain a bead and
an anti-bead. This should slow or stop the evolution of $r$ from loop formation. 
This argument suggests a scaling value for $r$ of  $ 2 M_b/(\mu\ell)$, though the contingencies of a fully realized network will doubtlessly complicate this simple picture. Nonetheless, this mechanism
seems to us robust, and should enforce a lower limit on $r$ for 
cosmic necklaces that can form loops.

Note that if $r$ grows in the scaling regime, then loop formation will only enhance that effect and there will be no scaling
in $r$ during an epoch of rapid bead production. Hence we find that if $\alpha < 10^{-4}$ (from section \ref{revolution}), 
then $r$ will grow. If this is the case, then cosmological considerations will force
 $M$ to be huge so as to suppress bead formation (see next subsection).

On the other hand, if $\alpha > 10^{-4}$ and Hubble stretching is tending to spread beads apart, 
then we find that $r$ scales to the following range of values (taking $\alpha \sim 0.1$ as an upper bound):
\begin{align}
10 M^{3/2} \frac{\sqrt{2g_s b}}{3\pi^{3/2}} \frac{H}{m_s} < r < 10^4 M^{3/2} \frac{\sqrt{2g_s b}}{3\pi^{3/2}} \frac{H}{m_s} \; .
\end{align}
Interestingly, the value of $r$ grows for large values of $M$.  Beads are potentially observable 
(though our analysis becomes less trustworthy) for values of order 1.  This would require very large values for $M$,
though, since $H/m_s$ is such a small number after inflation.  Unfortunately, since $H$ decreases with time, it seems unlikely that the beads will ever be observable. A value for $r$ of order 1 at the last scattering surface (with potential observable effects in the CMB) would mean a value of $r$ bigger than 1 at earlier times, where our analysis can no longer be trusted and
that very likely would lead to other cosmological problems.
Alternatively,  for the network to have $r$ of order 1 at the earliest 
time where we could hope for a lensing event to be observed -- at, say, a redshift of 1 -- would mean a value of $r$ of about $10^4$ at the surface of last scattering, leading to disastrous cosmological consequences. We also note that, dynamically, large values of $M$ actually 
suppress the creation of beads: it might thus be impossible for any dynamically-evolving network
to form enough beads to realize this scaling solution for $r$.

\subsection{Evolution in the Transient Regime}

It is much more difficult to analyze cosmic necklaces in the transient regime. 
In our network evolution, we have assumed that bead dynamics can be neglected. Although this is definitely true in certain cases (when $M$ is very large, for example), we might expect that certain initial conditions, together with certain values of the parameters, might lead to a large production of beads.

One thing we can say for certain: in the transient regime, $r$ initially \emph{grows}.   On
the one hand, $\alpha$ is expected to be of order $G\mu$ in the transient regime \cite{Vanchurin:2005pa, Bennett:1987vf, Allen:1990tv}, leading to more gravitational radiation production
per unit length. This should cause the beads to get closer to each other. 
Furthermore, we expect that the Hubble stretching is going to be less efficient before
we reach the scaling regime (since the string-gas pressure goes from 
positive to negative during the transient regime). Finally, a large initial density of strings
or a significant initial population of high quantum number states can lead to 
considerable bead production in the initial stages of the evolution, when higher physical
density will lead to far more string interactions and thus bead formation opportunities.

To get a feeling for how critical this situation is, we can do some simple approximate
calculations. To begin, we neglect bead production and loop formation. We will also 
assume that Hubble stretching is negligible. Taking $\kappa_s = 0$, we find
 $ r \sim t^{\kappa_g} = r_0 t^{4\pi G\mu/\alpha}$. Since we know that the scale 
 factor ``$a$" grows by at least 3 orders of magnitude before we reach scaling
 (which corresponds to 6 orders of magnitude in time during the radiation era),
 we can estimate how small $r_0$ needs to be assuming $r$ is of order $1$
 when we reach scaling. We find
\ba
r_0  \sim 10^{-6 (4\pi) G\mu/\alpha} \sim 10^{-10^{7}}\, ,
\ea 
where we have assumed the most extreme case:
$G\mu \sim 10^{-7}$ and $ \alpha \sim 10^{-12}$ in the transient regime. 
More realistic values for $\alpha$ will result in much larger and more reasonable
upper limits for $r_0$.
From this calculation, we see that we need  to start with essentially zero beads.  
Supposing that large quantum number $(p,q)$ states are initially populated,
then the only thing available for suppressing bead production is the tunneling coefficient.
  We get that $r \sim N(t) \sim e^{-M^2} \ln(t)$ from production. 
This means that $r_0$ grows like a log from production alone. Thus, we
 would need $e^{-M^2}$  to be of order $10^{-10^7}$, which implies $M \gg 10^{3.5}$.    
Again, this is the worst case scenario.  A better situation occurs if $\alpha \sim G\mu$. 
 In this case, we find $M \gg 10^{2}$. Let us emphasize that this analysis
 assumes that the high quantum states are present initially and that the only suppression
 is from tunneling.  On the other hand, we are still neglecting loop formation
as well as intercommutation, each of which increase $r$ and which will, thus, 
force us to require an even larger value for $M$.   Hence, if we have initial conditions in
which high quantum number states are populated,  we expect that $M$ needs to be 
even larger than our previous estimates for beads not to dominate. 
It is hard to accurately estimate this quantity, since we do not know how effective loop formation will be. 
We conjecture that $M$ of order of a thousand should be sufficient to keep the beads from dominating in the transient regime. In the less dangerous case -- when high quantum
number states are \emph{not} initially populated -- these requirements would be somewhat relaxed,
though a much more detailed analysis would be necessary to quantify by how much.

Finally, we expect that the number of beads produced depends on the 
initial density and spectrum of the string network the same way the final 
scaling density did.  This is very interesting, as it puts an indirect bound on quantities --
the initial string populations -- that are usually left totally unconstrained in standard string network
evolution. Indeed, cosmic string networks usually reach identical scaling solutions
for any initial density of cosmic strings.   On the contrary, we expect this not to be generically
true for cosmic necklaces. We do not believe this bi-modal scaling structure is an artifact 
of a poor statement of the problem, but in fact represents two physically distinct,
initial-condition-selected scaling solutions.

The Kibble mechanism predicts roughly one cosmic string per Hubble area at 
the time of formation.  For cosmic superstrings, there are some indications
 that this number may be much higher, of order of the string scale \cite{Barnaby:2004dz}.  
 Though such networks can reach scaling, the very large number
 of interactions they undergo at early times could lead to dangerously enhanced bead
 creation. Furthermore, it remains to be seen whether the same kind of analysis
 will suggest that the high-quantum-number states will be initially populated.
 Clearly this is yet another issue that needs to be explored further.  

In summary, we expect that the transient regime puts important 
constraints on both the value of M and the initial distribution of strings. However, we
cannot fully quantify these bounds without a more detailed analysis.

\vspace{-14 pt}
\subsection{Multi-tension String Networks with Beads}

In addition to small scale structure versus stretching, a binding, multi-tension
string network brings in other possible bead interactions. If an unzipping Y junction
meets a bead, for instance, it will ``grab" it. There will subsequently be three strings 
attached to the bead. Since the outgoing or ingoing 
charge will remain the same, we can always tell if it is a bead or an anti-bead. 
If two Y-junctions meet and the net bead number is non-zero, then one could create an 
X-junction. This process could in principle lead to junctions with any number of 
legs (perhaps with strange gravitational lensing properties cf. \cite{Shlaer:2005ry}). 
However, for reasons similar to those given above, we expect the formation of such bead-bound
junctions to be rare.  Furthermore any junctions with more than three strings attached
 to it should be unstable to decay into multiple Y-junctions.

\section{Conclusion}

Multi-tension cosmic string networks that are formed and live in phenomenologically viable warped throats
exhibit cosmologically safe scaling behavior under the following assumptions: (1) string binding 
proceeds as in Ref. \cite{Tye:2005fn} (2) Non-coprime states sometimes dynamically decay, 
even though they have a nonzero binding energy. We find
$\Omega_{strings} \sim 40 (70)   (8\pi/3) G \mu_F $ for $g_s = 1.0 (0.5)$ in our model, assuming
the simplest initial conditions.
The warped geometry causes these strings to have a tension that is periodic in $F$-string charge, $p$. 
We find that this property causes the results of network evolution to be
dependent on initial conditions: if strings with $p<M/2$ are the only
states initially populated, we find the above-stated results; but if we start with even a small
initial population of $p>M/2$ strings, the scaling value of $\Omega$ is more than doubled.

Additionally, the formation of any bound states that ``pass through" the maximum value for $p$
leads to the formation of beads on the string. This causes the string network to become decorated with
D-brane beads: a cosmic supernecklace. These beads, or monopoles,
 are cosmologically safe if the average loop size is larger than $10^3 G\mu t$ in the scaling regime,
so long as their production in the transient regime is sufficiently suppressed. Including annihilation in the analysis could potentially relax this constraint even further. 
In the transient regime, we have further assumed that the number of beads 
is small enough that it does not affect the network evolution. 
This condition is dependent on the mass of the beads (related to the flux number $M$) as well as the initial distribution of string populations. We expect these cosmological constraints
to give direct  bounds on $M$, though more detailed analysis will be needed to quantify those bounds.
Finally, we argued that the inclusion of loop formation could lead to a scaling solution of the interbead distance.  It might therefore be possible for beads to have observable effects, though
realizing such a scenario appears to require fine-tuning.
In any event, the beads' potential for creating distinctive observational
phenomena should be more thoroughly explored.

Looking forward, the methods discussed in this paper should allow us to study even
more general multi-tension string networks. It was conjectured in Ref. \cite{Firouzjahi:2006vp}
 that the general (p,q) string tension in a flux background should be given by
\begin{align}\label{conjecture}
T_{p,q} \simeq  m_s^2\sqrt{  g(q)^2 + f(p)^2}.
\end{align}
With appropriate parametrization, it should be possible to study the
behavior of networks with general $g(q)$ and $f(q)$ using the techniques outlined here.
As cosmological tests of cosmic strings become more stringent, such schemes may allow us to place 
useful limits on the structure of the string tension spectrum. This will give
string theory model builders guidance in constructing cosmologically viable, 
phenomenologically distinctive theories for cosmic string formation.  
While the ``square-root"
tension spectrum may not be a distinctive signature of string theory, it is rather hard to reproduce it in standard effective field theory framework (see Ref. \cite{Jackson:2006qc} for an attempt).  Saffin
has proposed an effective field theory model with two $U(1)$ fields to mimic (p,q) binding
\cite{Saffin:2005cs}, an approach anticipated by Carroll and Trodden 
\cite{Carroll:1997pz}.  Extending such models to produce periodic tension formulae is the
next challenge.

Cosmic superstrings remain one of the most compelling potential observational windows
into physics at the string scale. They appear to be cosmologically safe; they can
be created and maintained as metastable in realistic string theory models; and, as we have
shown, their phenomenology is rich and varied. If they do exist and can be observed, they 
might provide unparalleled observational access to the physics of string theory.

\acknowledgements{We thank Melanie Becker, Aaron Bergman, Jose Blanco-Pillado, Jason Kumar,  Ken Olum, Joe Polchinski, Sarah Shandera, Benjamin Shlaer,  
Henry Tye, Justin F. Vazquez-Poritz, Steve Thomas and Ira Wasserman for comments
and conversations. L.L. is supported by NSF grant PHY-0555575 at the Institute for 
Fundamental Physics at Texas A \& M University and thanks the Perimeter Institute for 
hospitality during the completion of this work. The work of M. W. at the Perimeter Institute is
supported in part by the Government of Canada through NSERC and by the Province of
Ontario through MEDT. This work was made possible by the facilities of the Shared Hierarchical Academic Research Computing Network (SHARCNET:www.sharcnet.ca).}

\bibliographystyle{h-physrev}

\end{document}